\documentclass[sigconf]{acmart}
\AtBeginDocument{%
  \providecommand\BibTeX{{%
    \normalfont B\kern-0.5em{\scshape i\kern-0.25em b}\kern-0.8em\TeX}}}

\setcopyright{acmlicensed}
\copyrightyear{2024}
\acmYear{2024}
\acmDOI{XXXXXXX.XXXXXXX}
\usepackage{makecell}
\acmConference[]{}
\long\def\comment#1{}

\begin{document}

\title{The Illusion of Anonymity: Uncovering the Impact of User Actions on Privacy in Web3 Social Ecosystems}

\author{Bin Wang}
\affiliation{%
  \institution{Beijing Jiaotong University}%
  \city{Beijing}%
  \country{China}%
}

\author{Tianjian Liu}
\affiliation{%
  \institution{Beijing Jiaotong University}%
  \city{Beijing}%
  \country{China}%
}

\author{Wenqi Wang}
\affiliation{%
  \institution{Beijing Jiaotong University}%
  \city{Beijing}%
  \country{China}%
}

\author{Yuan Weng}
\affiliation{%
  \institution{Beijing Jiaotong University}%
  \city{Beijing}%
  \country{China}%
}

\author{Chao Li}
\affiliation{%
  \institution{Beijing Jiaotong University}%
  \city{Beijing}%
  \country{China}%
}

\author{Guangquan Xu}
\affiliation{%
  \institution{Tianjin University}%
  \city{Tianjin}%
  \country{China}%
}

\author{Meng Shen}
\affiliation{%
  \institution{Beijing Institute of Technology}%
  \city{Beijing}%
  \country{China}%
}

\author{Sencun Zhu}
\affiliation{%
  \institution{Pennsylvania State University}%
  \city{University Park}%
  \country{PA, USA}%
}

\author{Wei Wang*}
\affiliation{%
  \institution{Beijing Jiaotong University}%
  \city{Beijing}%
  \country{China}%
}
\email{wangwei1@bjtu.edu.cn}

\begin{abstract}
  The rise of Web3 social ecosystems signifies the dawn of a new chapter in digital interaction, offering significant prospects for user engagement and financial advancement. Nonetheless, this progress is shadowed by potential privacy concessions, especially as these platforms frequently merge with existing Web2.0 social media accounts, amplifying data privacy risks for users. 
  
  In this study, we investigate the nuanced dynamics between user engagement on Web3 social platforms and the consequent privacy concerns. We scrutinize the widespread phenomenon of fabricated activities, which encompasses the establishment of bogus accounts aimed at mimicking popularity and the deliberate distortion of social interactions by some individuals to gain financial rewards. Such deceptive maneuvers not only distort the true measure of the active user base but also amplify privacy threats for all members of the user community. We also find that, notwithstanding their attempts to limit social exposure, users remain entangled in privacy vulnerabilities. The actions of those highly engaged users, albeit often a minority group, can inadvertently breach the privacy of the larger collective. 
  
  By casting light on the delicate interplay between user engagement, financial motives, and privacy issues, we offer a comprehensive examination of the intrinsic challenges and hazards present in the Web3 social milieu. We highlight the urgent need for more stringent privacy measures and ethical protocols to navigate the complex web of social exchanges and financial ambitions in the rapidly evolving Web3.
\end{abstract}





\maketitle

\section{Introduction}
Over the past decade, social media platforms have played a crucial role in reshaping social interactions, becoming an integral component of daily life. Since the inception of pioneers like MySpace \cite{jones2008whose} and Facebook \cite{wilson2012review}, these platforms have revolutionized communication experiences and have opened up numerous opportunities for businesses to engage with consumers. 
However, with the rise of Web3—an iteration of the Internet that incorporates concepts such as decentralization, blockchain technologies, and token-based economics—established social media platforms are now compelled to reassess and adapt their operational models to remain relevant in an increasingly dynamic and rapidly evolving digital landscape. Web3 emphasizes user sovereignty and data privacy. It typically relies on blockchain's distributed data storage and state consensus to ensure users can control their own data, rendering transactions and interactions transparent and trustworthy. The token of Web3 serves as a powerful incentive, rewarding users and contributors for their participation and investment in the ecosystem.



The foundational concept of a Web3 social network lies in establishing a decentralized, verifiable social graph, targeting the prevalent issues of platform fragmentation and monopolization of user data seen in existing social frameworks. Web3 ushers in a user-focused decentralized social graph protocol, facilitating the generation, modification, acquisition, and confirmation of social graph data. Concurrently, Web3 cultivates a distinctive feedback loop within the data network, granting users control over their own data.


With the significant influx of users into the Web3, a variety of demands have arisen, leading to the creation of numerous social applications including Steemit, TOMO, Friend3, Cipher.rip, Post.Tech, and friend.tech. In response to the Web3 community's call for unrestricted social interaction, Steemit, a blockchain-powered Web3 blog and social media platform, was launched in 2016. This platform rewards users for generating and curating content through token compensation. Following Steemit, platforms like Minds and Mirror have emerged, enabling users to produce content and monetize their contributions. Many of these platforms provide features for linking social accounts, allowing users to connect to their Web2.0 social media profiles, such as accounts in Twitter (rebranded as X) and Facebook, to ease the transition to Web3.


\begin{figure}[h]
  \centering
  \includegraphics[width=0.7\linewidth]{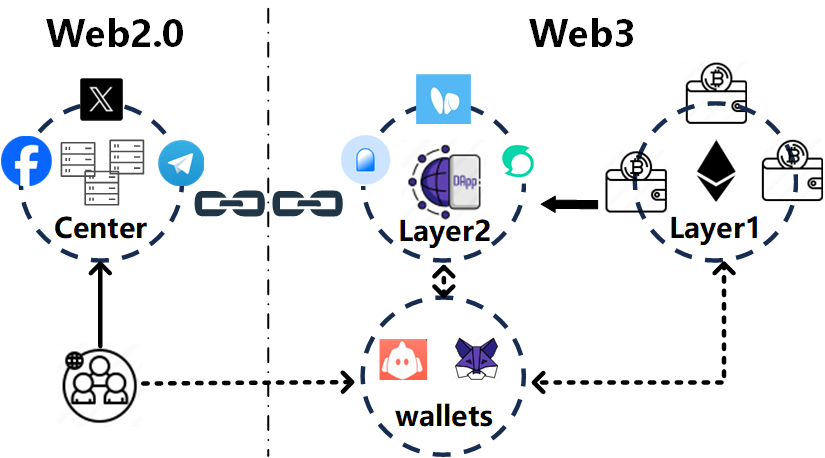}
  \caption{Web3 social platforms.}
  \Description{Web3 social platforms.}
  \label{Web3link}
\end{figure}

As depicted in Figure \ref{Web3link}, most modern Web3 social applications are built on the Ethereum blockchain \cite{buterin2013ethereum}. Initially, each user must create a wallet account on the layer 1 Ethereum network. When registering an account in a layer 2 decentralized social application, the application sends a request to associate with the Ethereum account of the user through the wallet. This association with the layer 1 Ethereum account acts primarily as a safeguard against Sybil attacks \cite{douceur2002sybil}, which is enforced by validating the balance of accounts. However, the transaction data generated during the account creation process poses significant privacy risks for users.


As decentralized social platforms attract an increasing number of users, the complexity of user interactions on layer 2 expands, creating an intricate social network. These interactions have significant privacy implications, and the data linking users is valuable to both users and network owners. Moreover, the ability for layer 2 to connect with layer 1 allows various stakeholders, such as service providers \cite{rehman2019cloud}, data analysts \cite{kalodner2020blocksci}, and government agencies, to create and examine social graphs on layer 1. This activity threatens the anonymity that is a hallmark of blockchain technology and which some users wish to maintain. For example, the friend.tech initiative by Twitter \cite{liu2023tokenomic} connects traditional Web2.0 Twitter accounts with the decentralized Web3 environment. Given that Twitter contains extensive user data, such as geolocation, personal images, and behavioral patterns \cite{wu2011says}, transferring these social connections to Web3 could significantly undermine the anonymity that blockchain systems are known for.


During our research, we uncovered that Web3 social platforms encounter considerable risks of privacy leakage, particularly during the data exchange processes between layer 2 and layer 1, as well as in the integration between Web3 and legacy Web2.0 platforms.

\textbf{Cross-Layer Privacy Leakage.} Transactions between layer 2 and layer 1 have the potential to compromise privacy, as user actions on layer 2 can be associated with accounts on layer 1. Particularly, the behavior associated with linking multiple addresses is observed when users engage in activities to capitalize on arbitrage opportunities. These linkages could reveal that layer 1 addresses, which appear unconnected, might actually be owned by a single user.

\textbf{User Identity De-anonymization.} Owing to insufficient privacy protection measures on Web2.0 platforms, associating the accounts across layers could unintentionally subject Web3 data of users to the surveillance and analysis that is common in Web2.0 environments. Such exposure may, to a degree, uncover the real-world identities of individuals behind layer 1 accounts.

In this study, we explore how social networks on layer 2 influence the anonymity of blockchain accounts on layer 1. We begin by outlining the characteristics of a Web3 social platform ( detailed in Section 2). Our analysis, covering the period from its inception to the April 2024, uncovers a significant portion of fraudulent activities. This finding prompted us to investigate the consequences of layer 2 user behaviors on the anonymity of layer 1 accounts. The complex interplay between social interactions and financial transactions led us to examine a broad transaction network and analyze the fundamental records of the blockchain (further elaborated in Section 3). We discover that the impact on user privacy is substantial, and we provide a detailed account of the extent and durability of this impact. Furthermore, we investigate how high-follower users affect the privacy of those with fewer followers, examining how the activity levels of influential platform users can jeopardize the privacy of less active members (as analyzed in Section 4).

Our key contributions are as follows:
\begin{itemize}
\item To the best of our knowledge, this study presents the first comprehensive examination of link privacy within the context of Web3 social networks, by analyzing an extensive dataset comprising 152,600 users and 621,301 transactions.

\item Our study delves into the emerging domain of Web3 social applications by using friend.tech, a platform initiated by Twitter, as a case study. We analyze patterns of user engagement and identify specific exploitative financial activities, such as wash trading and bonus hunting, that are prevalent among users.

\item Our research establishes that inter-user relationships on layer 2 social networks significantly affect the anonymity of layer 1 users. Our quantitative analysis reveals that extended interactions on these platforms result in a greater erosion of anonymity. Furthermore, we find that the activities of highly engaged users on platforms like friend.tech can potentially compromise the privacy of less active members, raising important questions about the extent to which these networks challenge the core promise of anonymity inherent in blockchain technology.
\end{itemize}

\section{BACKGROUND}
\subsection{Web3 Social Platforms}
Web3 social platforms signify a significant advancement in digital interactions, leveraging DApps across various blockchain networks to augment user independence and participation. These platforms enable users to tokenize diverse facets of their online activities and interactions, pioneering a novel method for quantifying social influence.


To address scalability and efficiency, developers often implement a layer 2 sidechain (also called \textit{Base chain}) that runs in tandem with the primary blockchain, or layer 1 chain. This strategy involves transferring platform data onto the Base chain, which effectively reduces transaction costs, commonly referred to as ``gas fees'', and enhances the overall performance of the services offered. This innovation is pivotal in attracting a diverse user base and spurring the creation of a multitude of DApps, each designed to cater to the unique needs and interests of users within the Web3 space.

On Web3 platforms, the absence of centralized incentive structures typically motivates social activities through the issuance and exchange of mintable tokens. The economic worth of these tokens arises organically from the collective choices of the user base, granting token holders the opportunity to profit from trading operations.

\textbf{ETH.} is the native cryptocurrency of Ethereum. To augment the functionalities of Ethereum, the Base chain serves as a layer-2, application-focused blockchain. This layered design guarantees seamless interoperability of ETH between the foundational and application layers.

\textbf{Transaction.} A key characteristic of transactions on Web3 social platforms is the asset exchange between accounts. These platforms ensure that all user actions—be it transfers, purchases, or sales—are recorded as transactions on the Base chain. Each transaction's specifics are precisely documented in the ``method'' field.

Within the evolving landscape of Web3, social platforms such as TOMO, Friend3, Cipher.rip, Post.Tech, and Friend.Tech have emerged. As shown in Table \ref{table:platforms}, the Friend.Tech platform boasts the best performance in terms of user data and revenue among its peers\cite{IOSG01}. We will focus our research on Friend.Tech due to its comprehensive features and active community.

\begin{table}
\centering
\caption{The comparison of Web3 social platforms}
\label{table:platforms}
\begin{tabular}{ccccc}
\toprule
\textbf{Blockchain}& \textbf{Name}& \textbf{\makecell{Twitter\\ Followers}}& \textbf{\makecell{Total\\ Tx}}&\textbf{TVL}\\ 
\midrule
Base& Friend.tech& 151.4k& 11.1m&\$43.44m\\ 
Bitcoin& \makecell{New Bitcoin\\ City} & 21.6k& -&\$2.43m\\
 Linea& TOMO& 10.1k& 253k&\$1.34m\\
 BSC& Friend3& 31.1k& 141k&\$21k\\
 Arbitrum& Cipher.rip& 9.7k& 218.7k&\$362k\\
 Arbitrum& Post.Tech& 61.7k& 1.7m&\$187k\\
 Avalanche& Stars Arena& 30.3k& 1.2m&-\\
 Solana& Friendzy.gg& 5.6k& 114.7k&\$44k\\ 
Solana& Hub3.ee& 16.1k& 107.8k&\$29k\\ 
\bottomrule
\end{tabular}
\end{table}

\subsection{Friend.tech}

\begin{figure}[h]
  \centering
  \includegraphics[width=0.85\linewidth]{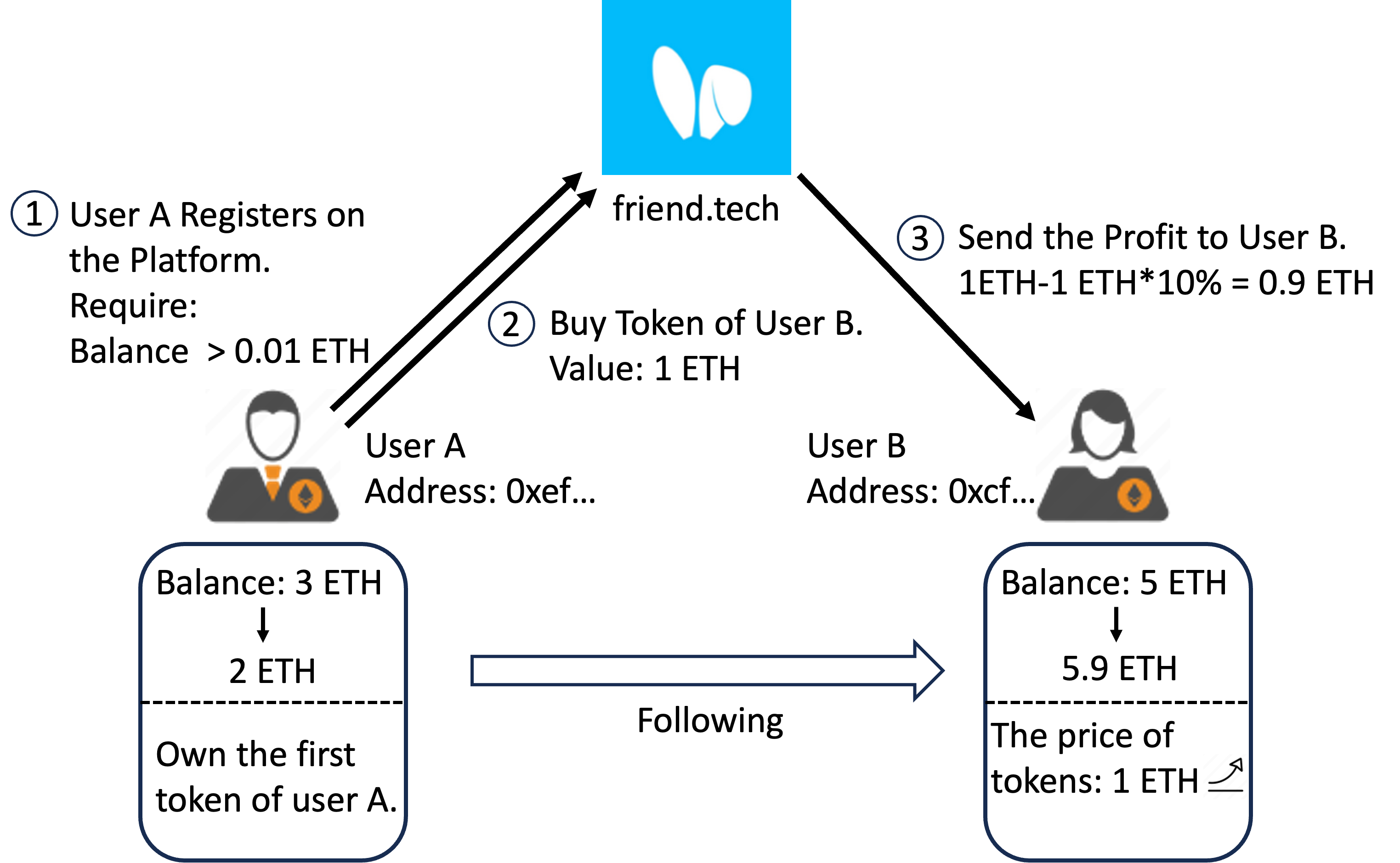}
  \caption{Friend.tech.}
  \Description{Friend.tech.}
  \label{friend_tech}
\end{figure}

\textbf{Overview.} Friend.tech is a DApp hosted on the Base chain and integrated with Twitter. This platform enables users to trade ``tokens'' associated with friend.tech users, streamlining transactions on the Base chain. The Friend.Tech platform allows users, such as User A, to tokenize themselves. As other users discover and purchase User A’s tokens, the selling price of these tokens increases. This strategy of converting popularity into profit is a key tactic used by social platforms to attract users. A smaller segment of highly activeusers, who often have the hiohest can be considered key opinion leaders (KOLs) within friend.tech wielding considerable influence. Investors in KOLs secure exclusive rights to direct communication with these influencers. Additionally, friend.tech facilitates the trading of tokens based on Twitter activities, granting shareholders privileged access to private communication channels. At its core, the platform capitalizes on the influence of KOLs by tokenizing their personal brands, offering an innovative method to monetize their social influence and establish a metric for social capital. 


\textbf{Users and social actions.} After granting permission to the application on Twitter and depositing 0.01 ETH, users can activate their profiles on the platform. Account creation initiates the setup of a specialized layer 2 account on the Base chain. This arrangement enables participants to partake in diverse social activities on friend.tech, utilizing the funds from their primary layer 1 accounts.



\textbf{Funds and transactions.} 
Once a user action is initiated, the Base chain records it as a transaction. Each transaction includes four essential components: {\itshape (i)} the input, specifying the address initiating the transfer; {\itshape (ii)} the output, denoting the recipient's address; {\itshape (iii)} the transacted amount; and  {\itshape (iv)}  the transaction type, such as token purchases or sales. This system enables users to establish multiple accounts, thereby providing opportunities to exploit arbitrage.

Figure \ref{friend_tech} presents a simplified model of user engagement on friend.tech. Consider user $A$, who, upon their inaugural entry into the platform, is subject to an account verification process requiring a minimum balance of 0.01 ETH. Successful verification grants user $A$ a set of initial tokens. Should user $A$, with an existing balance of 3 ETH, decide to purchase tokens from user $B$ for 1 ETH, the transaction would decrease the balance of user $A$ to 2 ETH post-purchase. Concurrently, the platform imposes a 10\% service charge, allocating 0.9 ETH to user $B$. This transaction not only affects the individual balances but also triggers a mechanism within the token economy model of platform. The acquisition of tokens by $A$ from $B$ leads to an appreciation in the market value of the tokens of $B$, illustrating the dynamic interplay between user transactions and token valuation.

Upon the completion of the transaction where user $A$ acquires tokens from user $B$, a social linkage is forged, exemplified by $A$ commencing to follow $B$. This event is indicative of the symbiosis between social interactions and the token-based economy on friend.tech, demonstrating how monetary exchanges are instrumental in constructing the social fabric of platform. The purchase transcends a mere financial exchange, evolving into a social transaction that potentially influences the connectivity and user behavior of network.

For ease of presentation, we define the essential terms used consistently throughout the paper.

\textbf{Action of Users.} This encompasses all types of social interactions performed by users on the platform, including the buying and selling of tokens and the creation of social graphs.

\textbf{Web3 Social Graph.} The social graph is constructed based on the interconnected follower relationships on the platform. In this graph, each node represents a user, and directed edges illustrate the follow relationships between them, indicating the direction of attention. The social graph can be considered a subgraph of the transaction network, reflecting the underlying social structure that often influences and is influenced by economic transactions.

\textbf{Token of Web3 Social Platforms.} Web3 social platforms are distinguished by their diverse token economic models. On the friend.tech platform, a user's token value hinges on the total volume of tokens they have issued, with larger issuances correlating to higher prices. This mechanism allows users to monetize their tokens for financial gain. However, some may manipulate this system by creating artificial scarcity to inflate profits. An example is the creation of multiple secondary accounts to buy out the token supply of a primary account, a tactic known as "wash trading." Such practices are generally discouraged as they can skew the true value and demand for tokens, thereby compromising the integrity of the economic framework.

\section{Datasets}
\subsection{Data Collection}

We developed a data extraction tool that utilizes the API services provided by friend.tech, specifically designed for gathering comprehensive user interaction data from the platform. Table \ref{tab:data} presents a summary of the data collected, including key metrics and trends.

\begin{table}
  \caption{Dataset description}
  \label{tab:data}
  \begin{tabular}{lll}
    \toprule
    Users&Transactions&Relationship\\
    \midrule
    id & Txhash & following \\
    address & Blockno & followers\\
    twitterUsername & UnixTimestamp & \\
    twitterName & DateTime& \\
    twitterPfpUrl & From& \\
    twitterUserId & To& \\
    lastOnline & ContractAddress& \\
    lastMessageTime & Value\_IN(ETH)& \\
    holderCount & Value\_OUT(ETH)& \\
    holdingCount & TxnFee(ETH)& \\
    watchlistCount & TxnFee(USD)& \\
    tokensupply & Method& \\
    displayPrice & Status& \\
    lifetimeFeesCollectedInWei & ErrCode& \\
  \bottomrule
\end{tabular}
\end{table}

\textbf{User Data.} The foundational phase of our research entailed curating an exhaustive roster of users featured on the platform. We compiled extensive profiles for each individual, capturing a spectrum of data including personal details, follower statistics, Twitter identifiers, and metrics on token ownership. Illustratively, user $A$ is assigned a distinct identifier, id, by the platform, with its account designated as address. We mapped the Twitter-related attributes of user $A$, such as twitterUsername, twitterName, twitterPfpUrl, and twitterUserId, to their profile. In our pursuit of dynamic user engagement patterns, we chronicled the activity timestamps—namely of users, their last online instance and most recent message, annotated as lastOnline and lastMessageTime, respectively. Furthermore, our analysis delved into the token economy, quantifying the holderCount as the volume of the tokens held of user $A$ across the user base, and the holdingCount as the tally of tokens in the possession of user $A$. The watchlistCount metric gauges the visibility and interest in the profile of user $A$, as indicated by the count of profile views. Conclusively, tokensupply quantifies the aggregate of the tokens available of user $A$, while displayPrice provides a real-time valuation of said tokens in the marketplace.

\textbf{Transaction Data.} On the friend.tech platform, all user interactions are recorded as transactions within the Base chain infrastructure. The dataset we harvested encompasses various elements of transactional data, including but not limited to the type of transaction, associated fees, and the amounts transferred in and out (Value\_IN and Value\_OUT). Our data collection spans the entirety of transactions disseminated by users on friend.tech. The transaction hash, denoted as Txhash, serves as a unique identifier for each transaction. The block number, labeled as Blockno, specifies the transaction's location within the blockchain. The UnixTimestamp provides a precise temporal marker of when the transaction was officially recorded on the blockchain. The ``From'' and ``To'' fields demarcate the sender and recipient of the transaction, respectively. Value\_IN and Value\_OUT quantify the funds received and disbursed in each transaction, respectively. The transaction fee, or TxnFee, reflects the network fee incurred. The ``Method'' field classifies the transaction type-such as transfers, purchases, or sales—highlighting the diverse array of social and financial exchanges possible within this digital ecosystem.

\textbf{User Relationship Data.} We assemble data related to followers of users and those they are following as outlined within their profiles. For instance, the following list of user $A$ encompasses users they opt to follow. Analyzing this list enables us to pinpoint users whose tokens user $A$ has acquired. On the flip side, the followers list comprises individuals who have chosen to follow user $A$, from which we can deduce users who have bought the tokens of user $A$. This gathered information aids in constructing a comprehensive social graph for each participant, incorporating important relationships such as mutual connections.

\textbf{Summary.} Our dataset covers the period from August 2023 to February 2024. Within this timeframe, we systematically sampled and documented 152,600 data points related to addresses and 621,301 transactions. These addresses encompass both individual user addresses and contract addresses. Table \ref{tab:data} details the additional information we gathered for user addresses linked to friend.tech, such as their associated Twitter accounts. The transaction data is divided into three categories: {\itshape (i)} fan token transactions executed by users on the layer 2 via the friend.tech platform; {\itshape (ii)} activities like transfers and contract interactions conducted on layer 2; and {\itshape (iii)} similar activities performed on the Ethereum blockchain.


\subsection{Data Overview}

\begin{figure}[h]
  \centering
  \includegraphics[width=\linewidth]{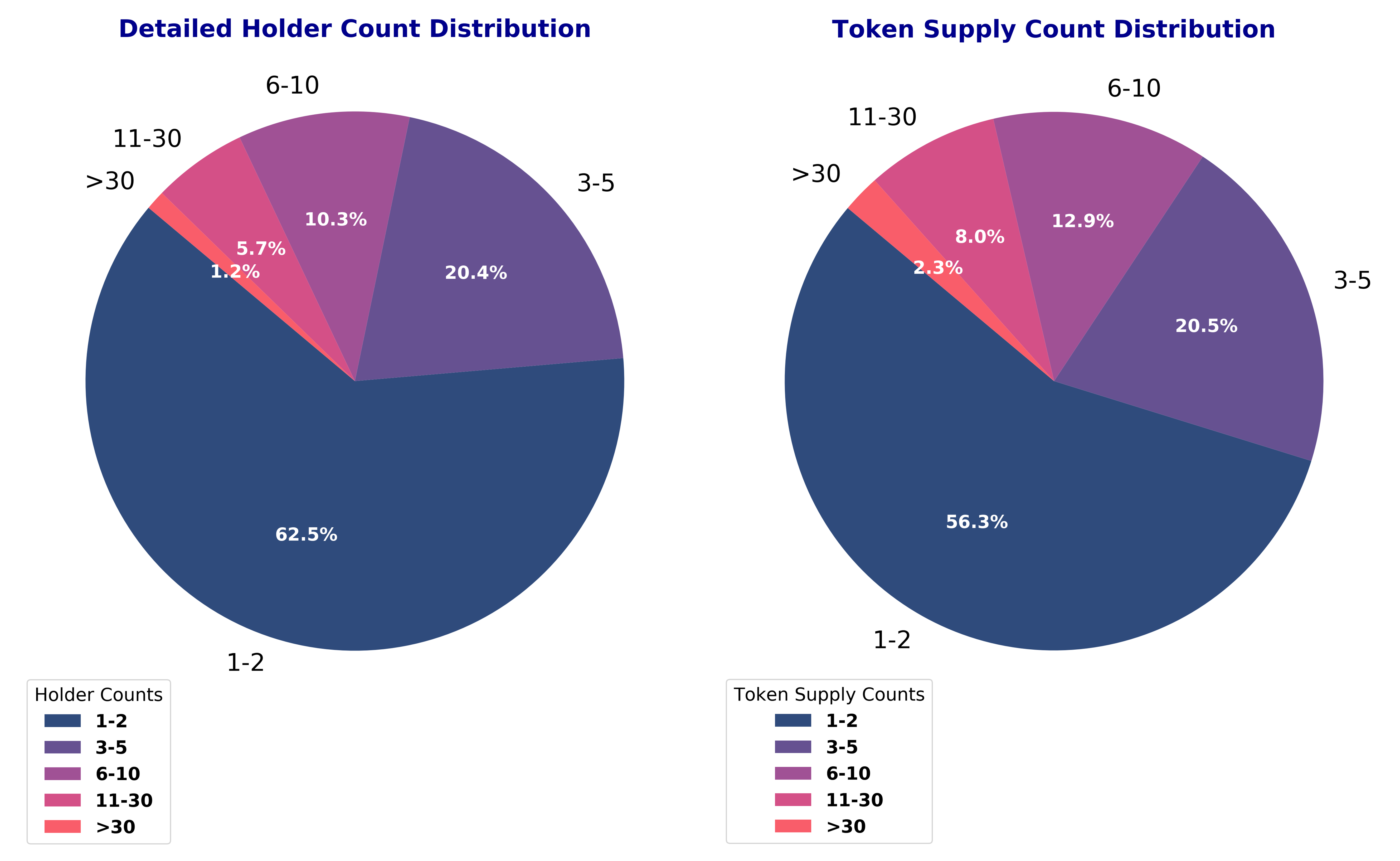}
  \caption{The distribution of tokens on friend.tech.}
  \Description{The distribution of tokens on friend.tech.}
  \label{SharesDistribution}
\end{figure}

\textbf{Distribution of Tokens.} We determine the preliminary economic of users standing by the number of tokens they hold, as illustrated in Figure \ref{SharesDistribution}. This quantity of tokens reflects the current market value of the tokens.

Our analysis shows that 62.5\% of users own less than 2 tokens, 20.4\% possess 3 to 5 tokens, and only 6.9\% have in excess of 10 tokens. This pattern suggests that most users either become inactive or engage only temporarily post-joining friend.tech, holding minimal tokens indicative of lower values. As a result, the average share price of platform is relatively modest. 

Token ownership patterns on friend.tech highlight significant variations in user engagement. A large portion of users demonstrate minimal investment, indicating low levels of commitment or interest, often reflected by reduced share values. Such trends could point to fundamental issues within the platform, potentially related to user retention, its value proposition, or strategies for engagement.

The data indicates that the token price of friend.tech exhibits a trend lower valuations overall, suggesting difficulties in deeply engaging users. The existence of a minority yet highly invested user group, possessing tokens of greater value, highlights opportunities for improvement in developing a more committed and faithful user community.


\begin{figure}[h]
  \centering
  \includegraphics[width=\linewidth]{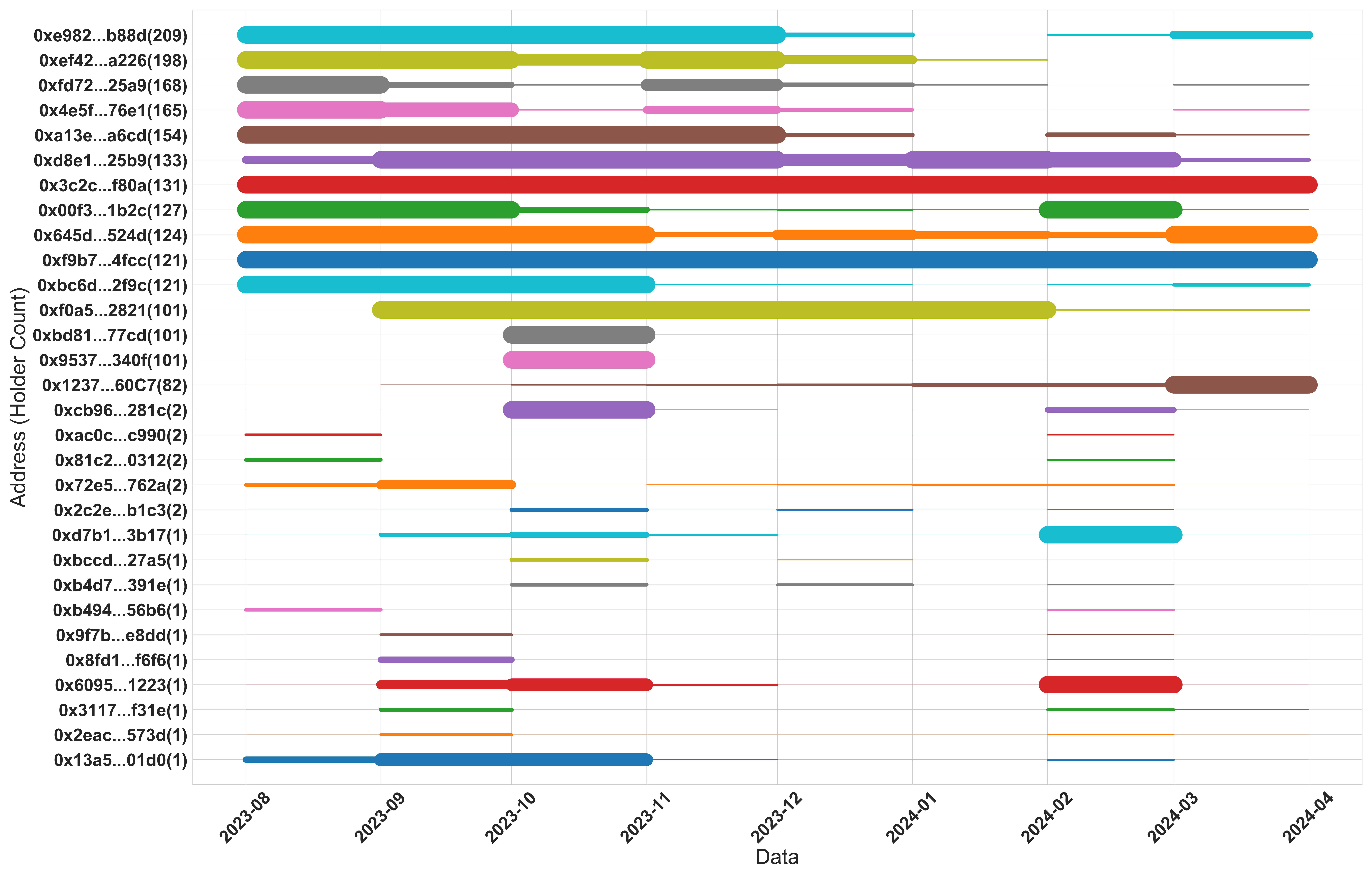}
  \caption{The user engagement with different tokens on friend.tech.}
  \Description{The user engagement with different tokens on friend.tech.}
  \label{UserActive_Shares}
\end{figure}

\textbf{User Engagement.} Our exploration of user engagement dynamics on the platform, particularly the relationship between user join dates and the accumulation of tokens (as illustrated in Figure \ref{UserActive_Shares}), revealed significant insights. We visualize the social activity of an account over time, typically on a monthly basis. If an account exhibits increased social activity in any given month—such as posting content, interacting with other users, or participating in platform-specific engagements—a more pronounced (thicker) horizontal line appears on the graph for that month, indicating heightened engagement. Conversely, if an account shows less social interaction in a specific month, the corresponding horizontal line on the graph is less pronounced (thinner), denoting diminished activity. If an account exhibits no social activity in a given month—meaning it did not post, interact, or engage socially—the graph shows no horizontal line for that month. 

We noted that users with a large number of tokens consistently engaged on the platform. Remarkably, one user, holding a share count exceeding that of more than 200 others, stands out. Moreover, among the top 10 shareholders, a glaring contrast emerges—the share gap between the top-ranked user and the fifteenth exceeds 100 tokens. This disparity highlights the substantial influence of KOLs on the platform, drawing in a dedicated following and distinguishing them as the primary participants of the platform.

\begin{figure}[h]
  \centering
  \includegraphics[width=\linewidth]{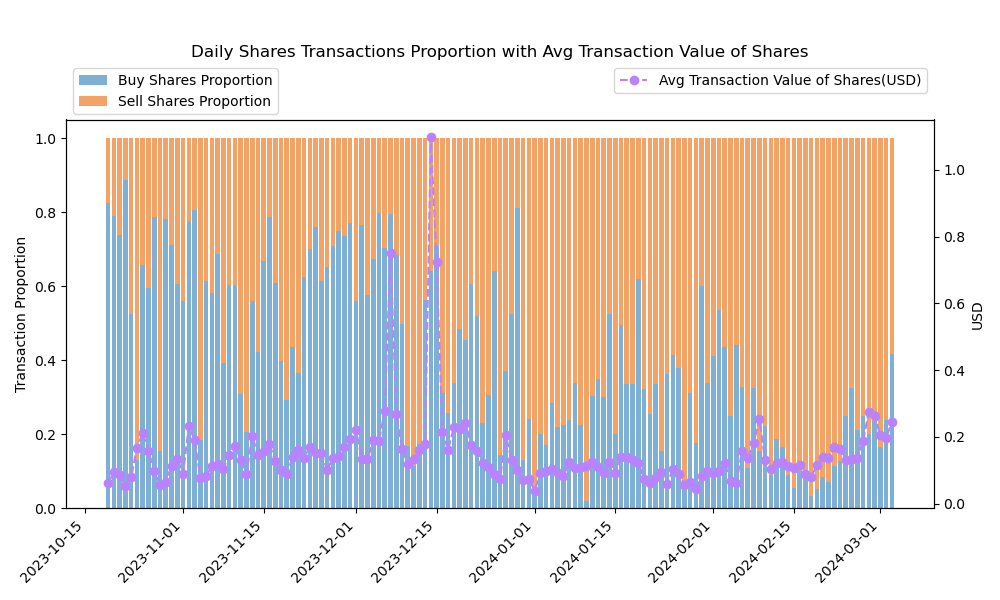}
  \caption{The daily transactions on friend.tech.}
  \Description{The daily transactions on friend.tech.}
  \label{DailyTransactions}
\end{figure}

Contrastingly, a comprehensive statistical analysis of users holding minimal numbers of tokens reveals a divergent trend. Unlike the KOLs, the bulk of these users entered the platform at subsequent stages, demonstrating a downturn in activities post four months. An exemplary case involves a user possessing merely 2 tokens, who maintained activity for less than a month. Subsequent to February 2024, a notable resurgence in activity was observed among users who had previously been inactive. Through our analysis of the friend.tech platform, it was found that significant updates, specifically the announcement of version 2, were shared on Twitter on January 29th, 2024, and again on February 24th, 2024. These announcements appear to correlate with the observed increases in user engagements. This trend suggests a prevalent indifference among users with fewer tokens towards potential economic losses, highlighting a significant deficiency in token economic model of friend.tech, which lacks sufficient incentives for the average user.


\begin{figure*}[h]
  \centering
  \includegraphics[width=\linewidth]{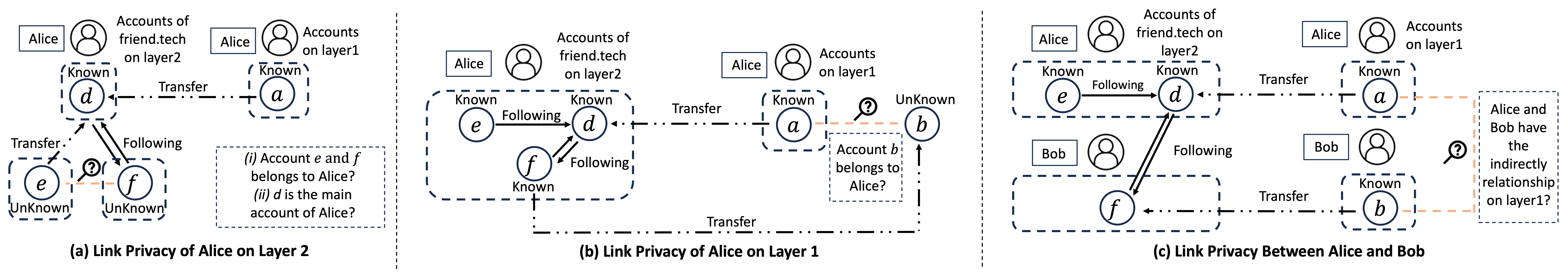}
  \caption{The link privacy of users.}
  \Description{The link privacy of users}
  \label{linkprivacy}
\end{figure*}

\textbf{The Trend of User Behaviors.} Our thorough analysis examined user social behaviors and the overall development of the ecosystem, including an assessment of share purchase price points as shown in Figure \ref{DailyTransactions}. We found that in the early stages of the platform, there was a strong inclination among users to buy tokens, with purchases accounting for an average of 70\%. However, after 2024, the proportion of users buying tokens saw a significant drop, averaging at 30\%. This indicates a marked decrease in user activity around five months after the inception of the platform, with more users opting to sell their tokens. Before 2024, there was a typical pattern of users accumulating tokens for an extended period before beginning to sell. Supporting this observation, our study found that a peak in share buying was often followed by a wave of intensified selling. This cycle can be attributed to token economy of friend.tech, where increased buying activity raises share values, leading some to sell at high prices for a profit, which in turn lowers share prices. Consequently, this attracts a group of users looking to take advantage of lower prices, creating a subset of users who benefit from these price fluctuations.

Closer analysis of the daily average transaction price for tokens showed that most were exchanged at lower prices, with KOLs engaging in trades less frequently. Before 2024, share prices generally followed an upward trajectory. However, this trend experienced a reversal due to two major sell-offs by KOLs in December 2023, which initiated a decline in share prices. This situation suggests that on Web3 social platforms, influential users hold the power to significantly influence the platform, potentially undermining its principle of decentralization.

\section{The Link Privacy of Users}


Using friend.tech as a case study, we explain the implications of layer 2 social platforms in Web3 on both individual and collective user privacy, as depicted in Figure \ref{linkprivacy}.

Considering potential user behavior, we have delineated three potential scenarios that may result in user link privacy leakage.

\textbf{ {\itshape(i)} Link Privacy of Users on Layer 2.} The ``Bonus Hunter'' phenomenon is a common practice that emerges during the early stages of some projects. Project teams implement promotional events offering free tokens or digital assets to boost user acquisition and retention. Many participants create multiple accounts to join these events and claim rewards. Their engagement typically wanes post-reward, leaving project teams discouraged by the prevalence of such behavior. On the friend.tech platform, bonus hunters typically adopt the strategy of generating multiple fake accounts. Once tokens are acquired from these accounts by automated bots or real users, the bonus hunters sell off their total assets for profit. To minimize transaction gas costs, they often merge their earnings into a single layer 2 account. The funds are then moved from this layer 2 account to a main layer 1 account.

In Figure~\ref{linkprivacy}(a), we present a scenario where a user, Alice, has created an account $d$ on a platform, linked to a specific Twitter account. We assume that unique ``twitterUserID'' identifiers correspond to individual Twitter accounts. On this platform, two other accounts, $e$ and $f$, exist, each associated with unique Twitter profiles. The identities behind accounts  and  are unknown to us, and they may potentially be auxiliary accounts belonging to Alice.
For instance, a financial transaction might be carried out from account  to account , where account  transfers funds to account  after realizing profits on the platform. Furthermore, accounts $d$ and $f$ follow each other, indicating a mutual interest. Interactions and transaction records between accounts $e$ and $f$ also exist. While accounts $e$ and $f$ are linked to separate Twitter profiles, our analysis might reveal that both accounts are actually controlled by Alice, with account $d$ serving as the primary interface for engaging with the platform's social environment. The presence of direct transactions between accounts $e$ and $d$ could suggest a relationship. If accounts $e$ and $f$ acted to economically benefit $d$ without a clear reciprocal service or product, it could imply that $e$ and $f$ are under Alice's control.

\textbf{{\itshape (ii)} Link Privacy of Users on Layer 1.} Suppose Alice operates account $d$ on a social platform, with accounts $d$, $e$, and $f$ each linked to a distinct Twitter profile. Utilizing the approach outlined in scenario {\itshape (i)}, we have identified hidden accounts $e$ and $f$ as likely belonging to Alice. Alice orchestrated a transfer from her primary layer 1 account $a$ to her secondary layer 2 account $d$, thereby activating account $d$ for registration and subsequent participation on the social platform. A financial transaction from account $f$ to account $b$ was recorded, signifying the transfer of accrued profits from the platform from $f$ to $b$. Additionally, accounts $d$ and $f$ engage in reciprocal ownership of the fan tokens of each other. Considering the transactional history between accounts $a$ and $b$, it stands to reason that account $b$ may also be associated with Alice on layer 1, analogous to account $a$.

\textbf{{\itshape (iii)} Link Privacy Between Users.} Broadening the context of scenario {\itshape (ii)} to encompass multiple users, we ascertain that Alice controls accounts $e$ and $d$, and Bob controls account $f$ on the platform. In the context of layer 1, Alice manages account $a$, and Bob manages account $b$. Observing that Alice and Bob possess the fan tokens of each other and share transactional history on layer 1, it becomes plausible to deduce a potential relationship between Alice and Bob on layer 1, derived from their engagement on the social platform. Correlating the dynamics of fan token exchanges with tangible financial transactions, we deduce that the nexus between Alice and Bob transcends mere social interaction on the platform, manifesting as a financial or strategic linkage on layer 1.

Users engage in repeated patterns of transactions or interactions that may inadvertently reveal their identity or relationship with other users. Over time, these patterns can be analyzed to infer connections and interactions between users, thus compromising their link privacy.

This subsection conducts a more in-depth analysis of user social graphs to ascertain the impact of layer 2 relationships on the link privacy of pseudonymous accounts in layer 1. Additionally, we propose a methodology to quantify the degree of link privacy degradation and assess its widespread occurrences.

\subsection{Privacy Leakage Analysis}

To better illustrate the analytical process of this paper, we will provide the following definitions.

\textbf{{\itshape Definition 1.} (Social Graph)} A social graph can be modeled as $G=\left(N,E\right)$, where $N=\left\lbrace{n}_1,{n}_2,\cdots,{n}_l\right\rbrace$ denotes the set of users on the platform who are interconnected socially, and $E\in N\times N$ denotes the set of edges representing connections between users. The set $E=\left\lbrace{e}^{transfer},{e}_{}^{sell},{e}_{}^{buy}\right\rbrace$ comprises the financial interactions among users, encompassing transfers and share transactions. The relational structure of $G$ is defined by an adjacency matrix $A$, in which ${A}_{i,j}=1$ indicates a connection between ${n}_i$ and ${n}_j$, and ${A}_{i,j}=0$ indicates no connection.

Real-world complex social graphs display a modular organization marked by communities of nodes with dense interconnections. The leakage of user privacy within this community structure may be conceptualized as follows: {\itshape (i)} interactions among nodes with concealed identities; {\itshape (ii)} interactions involving nodes with concealed identities and those with disclosed identities; and {\itshape (iii)} interactions among nodes with disclosed identities.



\begin{figure}[h]
  \centering
  \includegraphics[width=0.4\linewidth]{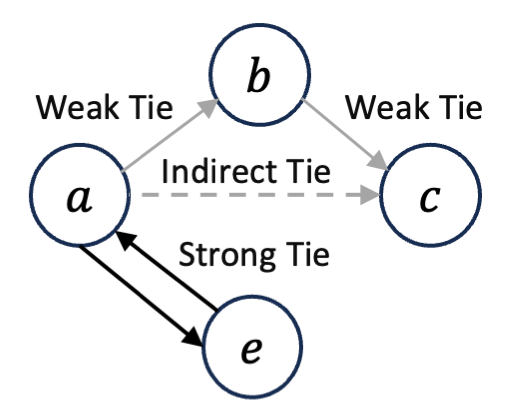}
  \caption{Strong tie between users.}
  \Description{Strong tie between users.}
  \label{Strong_tie}
\end{figure}

\textbf{{\itshape Definition 2.} (Strong Ties)} As illustrated in Figure \ref{Strong_tie}, the process of mutual following establishes relationships in three ways: {\itshape (i) Week Tie.} If user $a$ holds tokens of user $b$ but user $b$ does not hold any tokens of user $a$, a directed edge from $a$ to $b$ is established, denoted as ${A}_{a,b}^a={A}_{a,b}^b=1$ and ${A}_{b,a}^a={A}_{b,a}^b=0$. Similarly, if user $b$ holds tokens of user $c$ but user $c$ does not own any tokens of user $b$, a directed edge from $b$ to $c$ is established, denoted as ${A}_{b,c}^b=1$ and ${A}_{c,b}^c=0$. {\itshape (ii) Indirect Tie.} We denote the relationship between $a$ and $c$ as indirect tie. {\itshape (iii)} If users $a$ and $e$ mutually hold tokens of each other, a bidirectional connection between $a$ and $e$ is established, denoted as ${A}_{a,e}^a=1$ and ${A}_{e,a}^e=1$. We define the such relationship between $a$ and $e$ as a strong ties.

\textbf{{\itshape Definition 3.} (Privacy Leakage)}. Structural entropy measures the degree of node diversity in a network \cite{stantchev1986structural,Almog2019StructuralEM}. In our context, changes in node diversity, as indicated by alterations in the network, can act as a proxy for shifts in the social graph. The social graph of Ethereum is represented by $G$, comprising connected subgraphs $\{g_1$,$g_2$,$...$,$g_n\}$. With the advent of layer 2, $G$ transitions to $G'$, encompassing connected subgraphs $\{g'_1$,$g'_2$,$...$,$g'_m\}$. The probability vectors for $G$ and $G'$ are calculated as follows:

\begin{equation}
    \begin{split}
        V &= [\frac{c_1}{|N|}, \frac{c_2}{|N|}, ..., \frac{c_n}{|N|}] \\
        V' &= [\frac{c'_1}{|N'|}, \frac{c'_2}{|N'|}, ..., \frac{c'_m}{|N'|}]
    \end{split}
\end{equation}
where $c_i$ and $c'_i$ denote the number of nodes in subgraph $g_i$ and  $g'_i$respectively. $|N|$ and $|N'|$ represent the total number of nodes in $G$ and $G'$ respectively. Consequently, the structural entropy of $G$ can be computed as follows:
\begin{equation}
    \begin{split}
        H(V) &= -\sum_{i=1}^{n}\frac{c_i}{|N|}\log(\frac{c_i}{|N|})
    \end{split}
\end{equation}

The effect of layer 2 interconnections on structural entropy is quantifiable by the difference $H(V)-H(V')$. An example is provided to demonstrate our methodology.


\begin{figure}[h]
  \centering
  \includegraphics[width=0.9\linewidth]{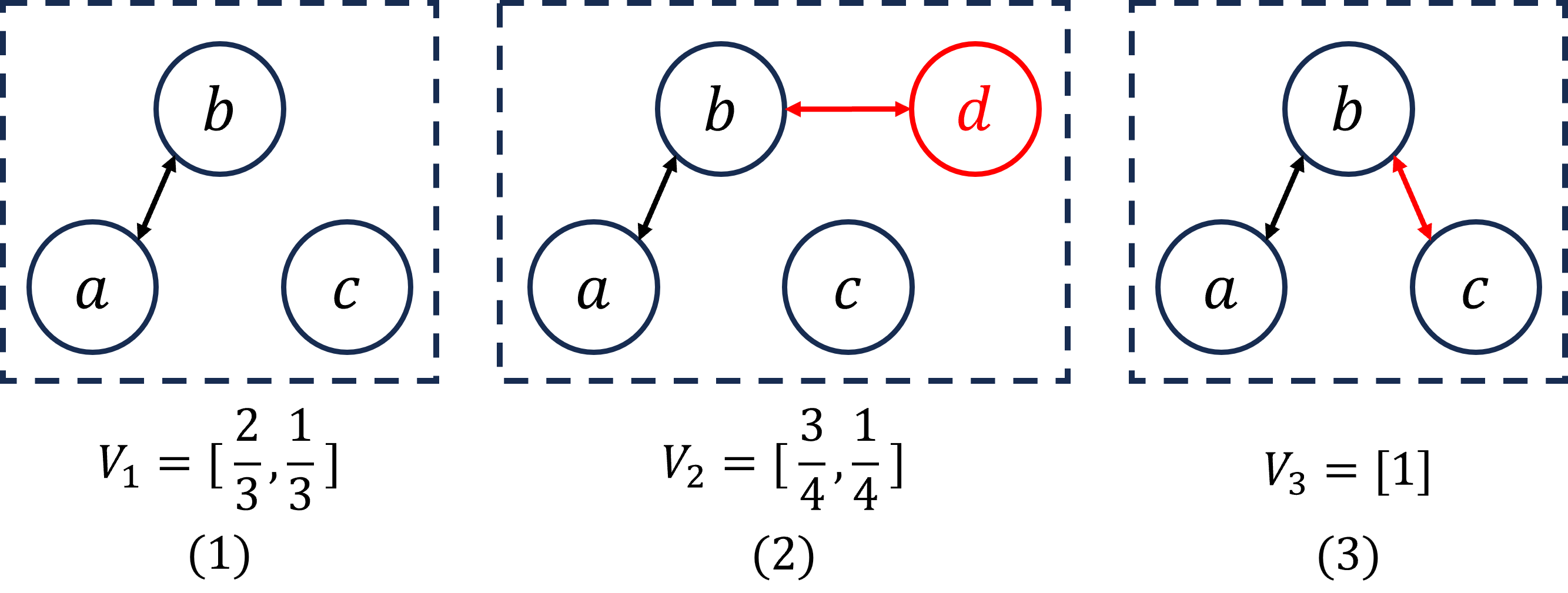}
  \caption{An example of structural entropy loss.}
  \label{explain}
\end{figure}

Consider the social graph on the Ethereum as depicted in Figure \ref{explain}. Layer 2 user interconnections influence structural entropy in several fundamental ways: the transition from (1) to (2) introduces a new edge linked to a new node, whereas (1) to (3) introduces a new edge connected to an existing node. Illustrating with the structural entropy of a weakly connected graph, the entropies $H(V_1)$, $H(V_2)$, $H(V_3)$ of states (1), (2), (3) are calculated as follows:
\begin{equation}
    \begin{split}
        H(V_1) &= -[\frac{2}{3}\log(\frac{2}{3}) + \frac{1}{3}\log(\frac{1}{3})] = 0.6365 \\
        H(V_2) &= -[\frac{3}{4}\log(\frac{3}{4}) + \frac{1}{4}\log(\frac{1}{4})] = 0.5623 \\
        H(V_3) &= -\log(1) = 0 \\
    \end{split}
\end{equation}
In both scenarios, the structural entropy has diminished, with a more pronounced decrease in the transition from (1) to (3). In (2), two existing nodes are connected, and in (3), only a relationship involving one existing node is introduced, which is intuitive.

In our social network, the edges that interconnect nodes are designated as relationships, which embody the inter-individual connections comprising the architecture of the network. 


\textbf{{\itshape Definition 4.} (Bonus Hunter)} We assume that account $a$ is the principal account controlled by the bonus hunter, with accounts $b$, $c$, and $d$ serving as subsidiary nodes instituted by the user. Subsequent to a sequence of social interactions, all benefits are eventually transferred to the main account. Consequently, our analysis is confined to the ``transfer'' and ``sell'' transactions between accounts, while other forms of social engagement are disregarded. The value transferred from account $b$ to $a$ is denoted as ${V}_{b,a}$. Similarly, the value transferred from account $c$ to $a$ and from account $d$ to $a$ are represented as ${V}_{c,a}$ and ${V}_{d,a}$, respectively. In an effort to curtail transaction gas expenditures, accounts $b$, $c$, and $d$ are expected to have transfer transactions with node $a$:

\begin{equation}
    {N}_b\cap{N}_c\cap{N}_d\ne\emptyset
\end{equation}
where ${N}_b$, ${N}_c$ and ${N}_d$ denote the set of users on the platform who are interconnected socially with account $b$, $c$ and $d$, respectively.

Before realizing their profits, bonus hunters typically sell off the tokens held in their fictitious accounts, following a consistent transaction sequence pattern:

\begin{equation}
\begin{aligned}
     \forall i\in\left\{b,c,d\right\rbrace,\exists {E}_{i}^{sell},{E}_{i}^{transfer} \subseteq {E}_{i},\\
     s.t.\left|{E}_i^{sell}\right|\gg\left|{E}_i^{transfer}\right|, {T}_{{E}_{sell}}^{i} \textless {T}_{{T}_{transfer}}^{i}
\end{aligned}
\end{equation}
where ${E}_{i}^{sell}$ and ${E}_{i}^{transfer}$ represent the transactions corresponding to the sale of tokens and the transfers executed by node $i$ respectively. ${T}_{{E}_{sell}}^{i}$ and ${T}_{{E}_{transfer}}^{i}$ indicate the timestamps of these transactions as recorded on the blockchain for selling and transferring, respectively.

Given the objective to pinpoint primary accounts maneuvered by users partaking in exploitative activities, namely ``bonus hunting'', the emphasis lies in tracing financial streams. A defining attribute of a primary account in such scenarios is its capacity to consolidate the highest profits and funds relative to other participating accounts. We consider the account with the maximum profit as potentially being the main account of a bonus hunter. By focusing on these main accounts, we can gain significant insights into link privacy. Consequently, these primary accounts might exhibit the following characteristics:
\begin{equation}
    a=\left\{i\in{N}_b\cap{N}_c\cap{N}_d\mid max {\sum}_{j\in\left\lbrace{j|A}_{j,i}^i=1\right\rbrace}{V_{i,j}}\right\}
\end{equation}
where $A^i$ is the adjacency matrix of node $i$. $V_{j,i}$ represents the values transferred from node $j$ to node $i$.

\subsection{Platform Characterization}
On friend.tech, users have the option to buy tokens from their peers. New users receive their personal tokens upon account creation. To delve into user engagement subtleties, we performed an analysis of the token ownership dynamics among users.

\textbf{User Activities.} In the friend.tech user ecosystem, a specific segment comprising 40.66\% users solely possesses self-issued tokens. These indicates significant issues regarding user engagement actively or operational functionalities. Such insights necessitate further research and corrective measures.

To highlight the significance of users with extensive shareholdings, we profiled 10 users, each holding over 50 tokens, as outlined in Table \ref{table:fans}. Furthermore, Table \ref{table:fancoinHolders} lists all individuals who own these tokens. We can see 46 users, each with 6 tokens from these 10 users, make up 26\% of the group following the elite. These insights indicate a notable concentration of users on the platform, marked by both direct and indirect relationships. The observed pattern delineates a pronounced KOL culture among the users. A marked concentration of token holdings within an echelon of elite users underscores the presence of a KOL-driven culture. These KOLs, through their significant token holdings, are poised to wield substantial sway within the community, influencing trends and opinions, and possibly affecting token valuations. Their influence is a testament to the social dynamics at play and the level of trust and credibility they have garnered within the ecosystem. The preferential trend towards token accumulation over selling among these elite users denotes a deep-seated confidence in the platform's potential and the perceived long-term value of the tokens. This behavior suggests a strategic investment approach, with users favoring the retention of their tokens in anticipation of future appreciation, as opposed to pursuing immediate gains through short-term trading.

\begin{table}
\centering
\caption{The Holder Count of 10 Users}
\label{table:fans}
\begin{tabular}{cc}
\toprule
Address & Holder Count \\
\midrule
0xa7d8...580c & 65 \\
0x8104...abd3 & 62 \\
0x9011...e526 & 62 \\
0xf0a1...532a & 60 \\
0x88e0...f56c & 60 \\
0xc2d1...ca33 & 60 \\
0x9d09...85fd & 60 \\
0x399d...0c0e & 60 \\
0xc4b2...ddb8 & 58 \\
0xa0db...e6f8 & 53 \\
\bottomrule
\end{tabular}
\end{table}

\begin{table}
\centering
\caption{Number of users holding the tokens of 10 users}
\label{table:fancoinHolders}
\begin{tabular}{lllllll}
\toprule
Token Number & 6 & 5 & 4 & 3 & 2 & 1 \\
\midrule
User Count & 46 & 2 & 1 & 27 & 28 & 73 \\
\bottomrule
\end{tabular}
\end{table}

\textbf{Real Users.} An analysis, as shown in Table \ref{table:Ties}, was conducted to determine the prevalence of strong ties among platform users. Among the top 30 users, 25.9\% of the connections were found to be strong, indicating a tendency among elite users to foster significant relationships. Further investigation into this segment of strong relationships revealed that 2.9\% are interconnected among the top users themselves. This observation indicates a trend where users are not merely retaining their tokens, but also investing in those held by their peers, implying a conviction in the potential expansion and profitability of their investments. The presence of robust connections often aligns with a heightened sense of trust and mutual valuation among the associated users. These relationships could potentially bolster cooperative actions and expedite the exchange of valuable data or resources. Conversely, the consolidation of influence and interlinked investments may present a market manipulation risk or create an aura of exclusivity, potentially impacting the platform's reputation among a wider user base negatively.

\begin{table}
\centering
\caption{Strong Ties of Different Users}
\label{table:Ties}
\begin{tabular}{cc}
\toprule
\textbf{Categories} & \textbf{Number of Strong Ties} \\ 
\midrule
Strong ties within the top 30 users & 144 \\ 
Strong ties related to the top 30 users & 4969 \\ 
Strong ties of all users & 19199 \\ 
\bottomrule
\end{tabular}
\end{table}

\subsection{Link Privacy of Users} 
The distinctive feature of friend.tech is its integration with layer 2. To complete registration, the newly created layer 2 accounts on friend.tech must hold a minimum balance of 0.01 ETH. This initiates a transaction, transferring funds from a existing account of a user, which may reside on either layer 1 or layer 2, to the new account. This provides a unique perspective for analyzing user transactions on the friend.tech platform and beyond.

\begin{figure*}[h]
    \centering
    \includegraphics[width=1\textwidth]{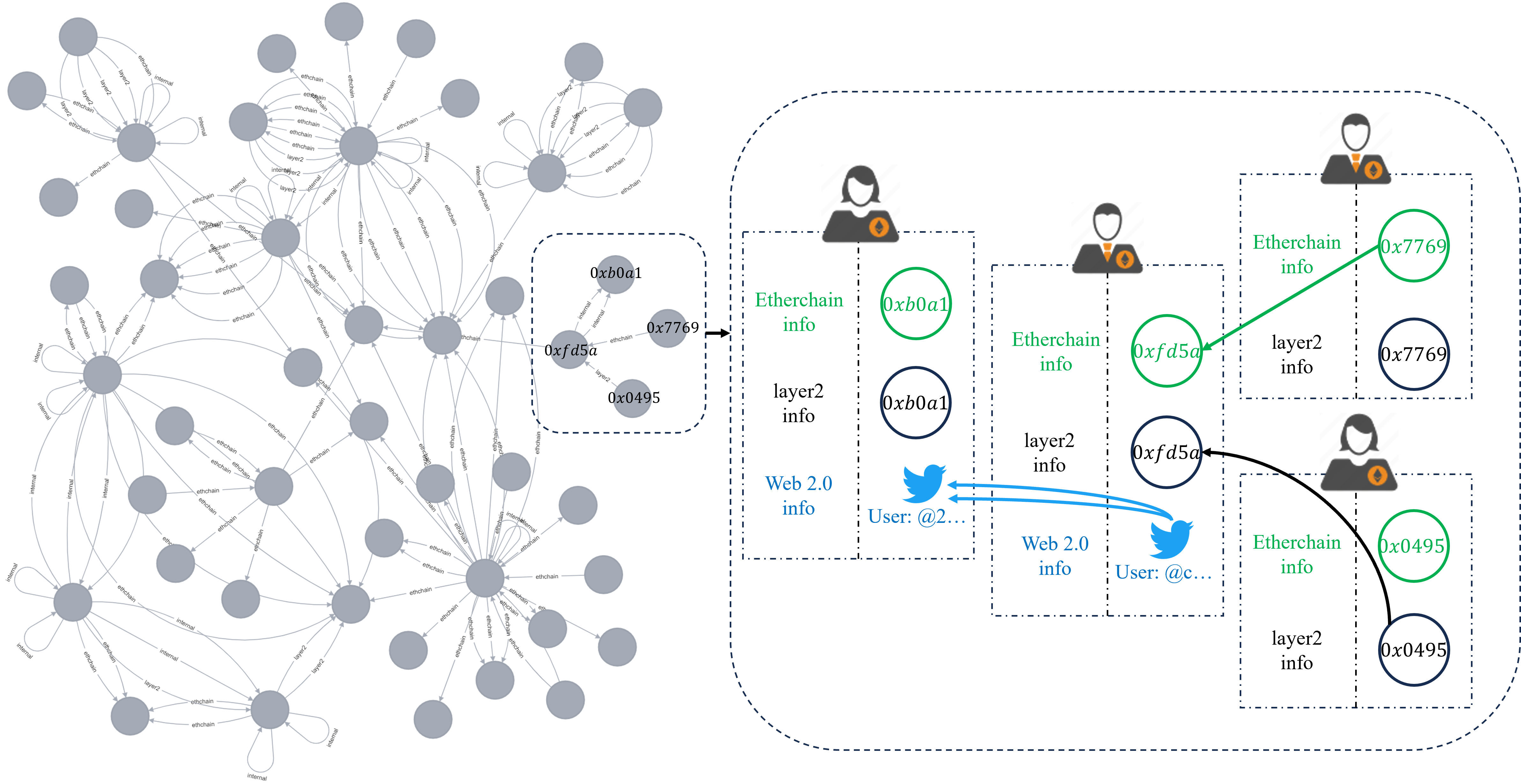}
    \caption{An example of link privacy leakage.}
    \label{fig:example}
\end{figure*}
In Figure \ref{fig:example}, we present a detailed illustration of link privacy challenges, highlighting the consequences of merging social networks, such as those from friend.tech, with blockchain infrastructures. Initially, of the four users illustrated, only two—denoted by their partial blockchain addresses, 0xfd5a and 0x7769—had a verifiable connection. However, the landscape transforms dramatically upon integrating their friend.tech contacts. Post-integration, the scenario unfolds in two significant ways: the Web2.0 personas of users 0xfd5a and 0x7769 are unequivocally revealed, and additionally, three new linkages materialize within this cohort. This case underscores the heightened risk to link privacy when conventional social networks are transposed onto a blockchain framework, potentially exposing users to unforeseen privacy breaches.

Our objective is to showcase the capacity of our technology in revealing user link privacy, thereby demonstrating the extent to which user anonymity is compromised on Web3 social platforms. Yet, we remain fully aware of the ethical implications and potential discontent that may stem from the excessive collection of personal data. In light of this, our strategy has been thoughtfully devised. We executed a small-scale experiment to evaluate the feasibility and effectiveness of our technology in correlating user information. It is crucial to emphasize that this was not a broad-scale data collection endeavor. Given the delicate nature of personal data, we ensured that only a minimal amount of information was collected, and strictly for research purposes.

Upon the completion of our research, we have committed to deleting all acquired sensitive data. This action reflects our dedication to user privacy and our adherence to ethical standards within our research. Moreover, it underlines our belief in the responsible use of technology, especially in the context of user data and privacy.

\begin{figure}[h]
  \centering
  \includegraphics[width=1\linewidth]{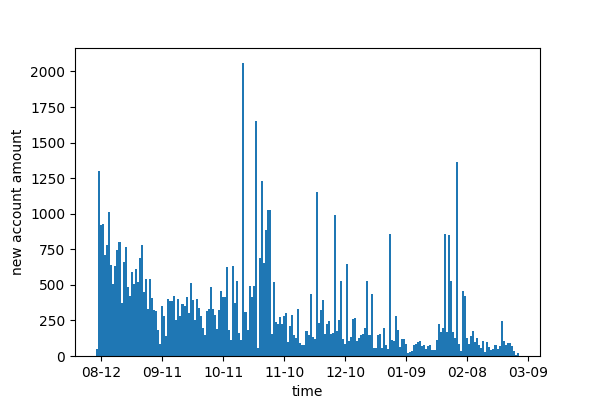}
  \caption{New users joined the platform.}
  \label{pic:time_static}
\end{figure}

Figure \ref{pic:time_static} depicts the inflow of new users to the platform. The trend shown in the figure reflects a steady decrease in the pace of new user registrations over time. A deceleration in the rate of new user acquisitions may indicate that the platform is nearing or has achieved a state of market saturation within its specified target demographic or geographic segment. With the maturation of platforms, the surge of initial exponential growth typically stabilizes, and a reduction in the influx of new users may signify a shift from an expansionary phase to a phase of stability and maturity. However, there are periods marked by sudden spikes in new users.

\begin{figure}[h]
  \centering
  \includegraphics[width=\linewidth]{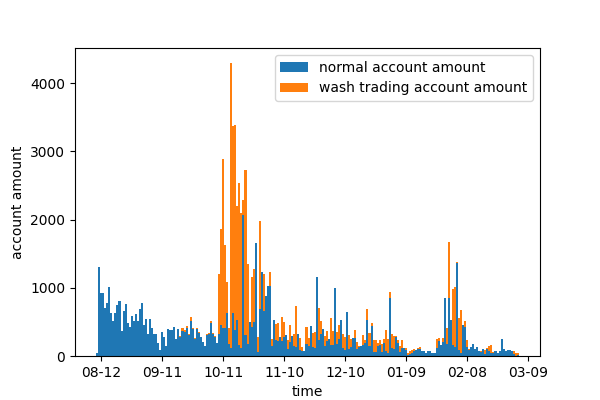}
  \caption{The washing tradings of friend.tech.}
  \Description{The washing tradings of friend.tech.}
  \label{timestamp_statistic_compare}
\end{figure}

\textbf{Washing Trading.} During our investigation, we observed that some users conducted transactions on layer 1, implying that they continued to be active on layer 1 through their newly registered accounts after registration. An interesting pattern emerged from October 9th to October 29th, 2023, where a high frequency of layer 2 accounts with a single ``Buy\_Share'' transaction was detected. These accounts represented a significant 97.42\% of the new users during this period, as shown in Figure \ref{timestamp_statistic_compare}. Intriguingly, these accounts were able to complete their registrations without an account balance exceeding 0.01 ETH, deviating from the standard transaction process of the platform. Despite our attempts, we were unable to obtain transactional data or additional information for these accounts on layer 1, suggesting their lack of activity. This evidence leads us to speculate that these accounts could have been created by the platform itself to simulate increased activity and encourage user participation, considering the inflexibility of the standard registration procedure for regular users.

For the detection and mitigation of wash trading, it is imperative to scrutinize transaction data for recurring patterns. Despite the pseudonymous nature of blockchain transactions, their public and traceable ledger enables the discernment of activities that suggest wash trading. Such patterns may unveil associations between accounts, thus posing a risk to user privacy. To circumvent detection, traders engaged in wash trading might resort to sophisticated strategies, including the orchestration of elaborate networks through the utilization of numerous interlinked accounts. Nevertheless, these subterfuges are not imperceptible and could inadvertently reveal linkages that threaten to erode privacy when deconstructed in the pursuit of identifying illicit actions.

\begin{figure}[h]
  \centering
  \includegraphics[width=1\linewidth]{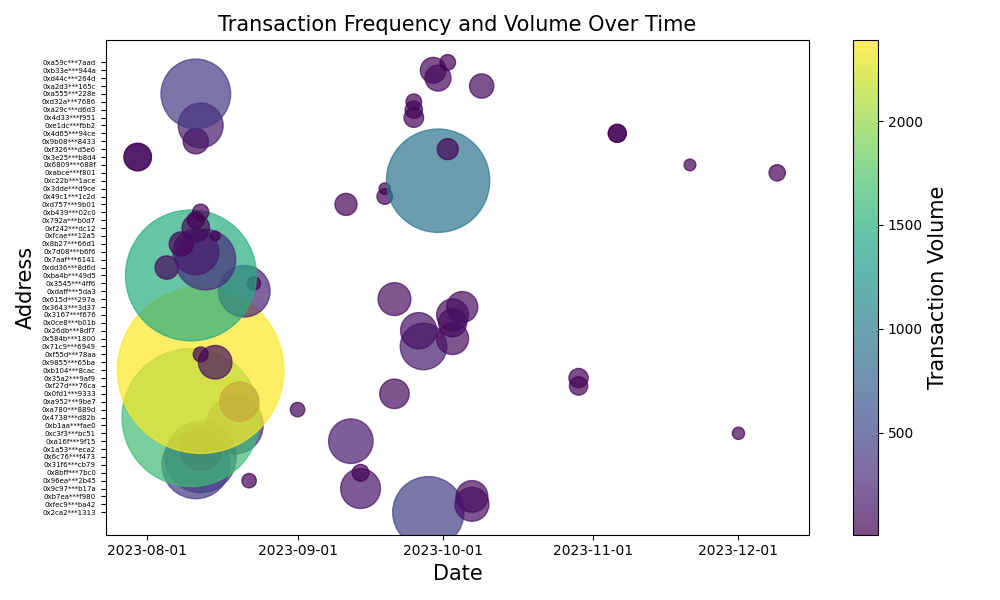}
  \caption{The details of bonus hunters over time.}
  \label{BonusHunter}
\end{figure}

\textbf{Bonus Hunters.} In-depth analysis of friend.tech revealed a focus on the exploitative user behavior known as "bonus hunting" and its temporal progression. As depicted in Figure \ref{BonusHunter}, bonus hunting was particularly prevalent during the platform's early stages, with a notable peak at its launch and a subsequent one in October 2023. Such spikes are typically linked to the platform's initial user acquisition strategies, which include offering attractive incentives. These incentives, during the embryonic stage, entice a significant influx of user registrations and activities aimed at capitalizing on the available bonuses. 

However, the data indicates a subsequent downtrend in bonus hunting, suggesting a reduction in authentic user engagement as well as a dip in the platform's aggregate activity. Analysis shows that 68.33\% bonus hunters conducted less than 100 transactions, indicating infrequent trading activities. We propose that this pattern is largely due to the 5\% fee levied on transactions, in addition to other associated costs, eroding the profit margins for bonus hunters. When transaction expenses surpass the anticipated bonus rewards, the motivation for these users to persist in such behavior markedly declines. Additionally, a natural corrective process may be influencing the ecosystem, where the commitment of long-term, genuine users supersedes the transient activities of bonus hunters. This authentic engagement likely results in lower attrition rates, as these users find value in the offerings of platform that extend beyond short-term incentives.


\subsection{Quantitative Assessment of Link Privacy Breaches}


The prevalence of transaction laundering on the platform has led to an inflation in the number of user accounts without corresponding inter-account interactions. Consequently, these accounts emerge as solitary nodes in the social graph, failing to enhance the link privacy for the general user base. In contrast, as the platform evolves towards stability and maturity, the user population stabilizes, resulting in a more static social graph architecture. Transaction laundering effectively narrows the anonymity set, as accounts implicated in such activities can be readily identified and segregated. Bonus hunting strategies, which entail the creation of multiple accounts to exploit incentive schemes, involve the circulation of funds across these accounts to fulfill the conditions for rewards. Such maneuvers establish explicit connections among numerous accounts, often characterized by consistent patterns like rapid, minor transfers across a vast network of accounts, thereby compromising the intended anonymity or pseudonymity of account linkages. Moving forward, we will undertake a quantitative assessment of link privacy breaches to elucidate the impact of these practices on link privacy over time, considering the fluctuations in the user base.

\textbf{Link Privacy.} Using Formula 1, we assessed the level of privacy leakage for users who joined friend.tech. We computed the metric $S\left(G\right)$ at different intervals to gauge this. Structural entropy reflects the unpredictability and intricacy of the connections of network. Elevated entropy indicates increased randomness and stronger privacy, as it complicates the tracking and forecasting of transaction patterns. On the other hand, reduced entropy signifies a simpler network structure, potentially making it easier to analyze and more vulnerable to privacy compromises. 

\begin{figure}[h]
  \centering
  \includegraphics[width=1\linewidth]{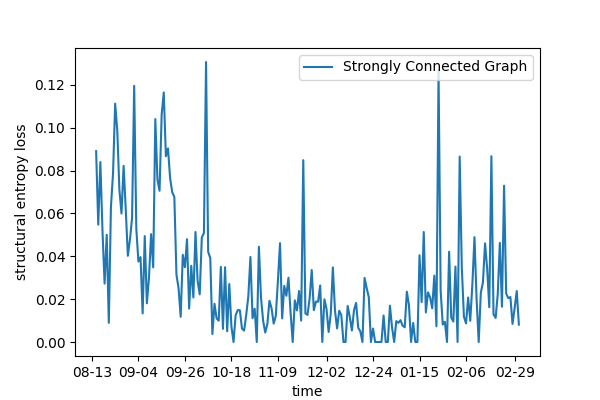}
  \caption{Daily structural entropy loss.}
  \label{pic:structural_entropy2_s}
\end{figure}

Figure \ref{pic:structural_entropy2_s} presents our preliminary analysis of the structural entropy changes in strongly connected graphs, measured at two-day intervals, starting from August 12th (Day 0). Each data point reveals the extent of privacy leakage from layer 2 user transactions from day $i$ to day $i+1$. During the peak activity phase between August and September, a significant drop in structural entropy was recorded, indicating increased privacy risks due to intense trading activities among users. As platform activity decreased, the rate of entropy loss also diminished. From January 2024 onwards, a slight increase in structural entropy loss was observed.

We deduce that during periods of high transaction frequency and user interaction, the complexity of transactions—and thus, the associated privacy risks—increased. This heightened activity may have generated more interconnected and predictable transaction patterns, susceptible to privacy breaches. Conversely, as user activity waned, the likelihood of privacy leaks decreased, leading to an increase in structural entropy and enhanced privacy protection.

The modest increase in entropy loss observed post-January may suggest a revitalization of platform activity or a fundamental alteration in the user transaction patterns, potentially due to the return of previous users or the rollout of new functionalities.

\begin{figure}[h]
  \centering
  \includegraphics[width=1\linewidth]{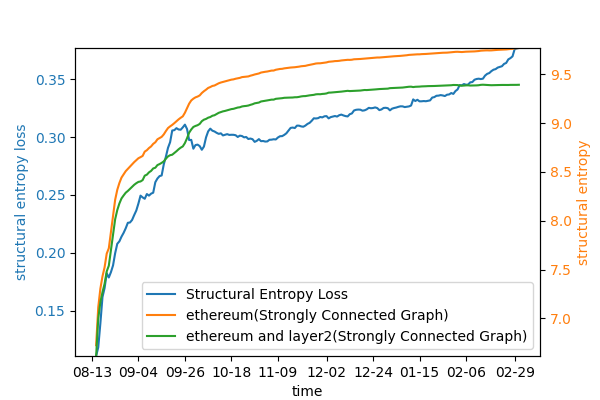}
  \caption{Accumulated structural entropy loss.}
  \label{pic:structural_entropy_s}
\end{figure}

Figure \ref{pic:structural_entropy_s} illustrates the trend of cumulative structural entropy loss beginning on August 12th, marked as Day 0, providing insight into the evolution of user privacy leakage within the layer 2 network. The data points on the graph denote the total privacy leakage from user transactions starting from Day 0 to the i-th day. Initially, the platform's nascent stage saw a marked increase in activity from new users, leading to a swift escalation in the cumulative loss of structural entropy. However, as the platform evolved and the influx of new users relative to interactions diminished, a notable deceleration in the rate of structural entropy loss became evident.

The relationship between the expansion of friend.tech and its subsequent effect on user privacy leakage presents a critical consideration. As friend.tech expands, it is reasonable to conjecture that privacy leakage may escalate disproportionately.

\begin{figure}[h]
  \centering
  \includegraphics[width=1\linewidth]{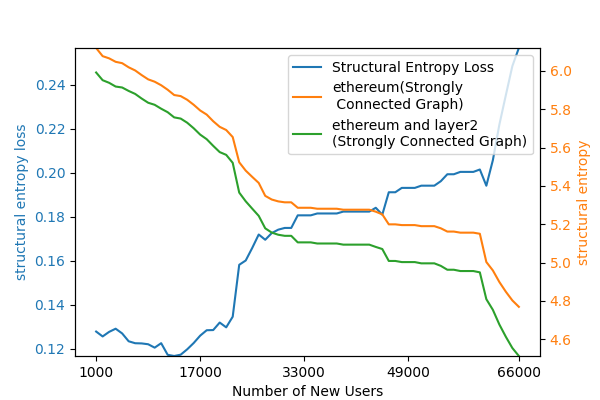}
  \caption{Link privacy of users with new users joined.}
  \label{pic:structural_entropy_index_s}
\end{figure}

The evidence presented in Figure \ref{pic:structural_entropy_index_s} substantiates a pronounced linear correlation between the influx of new users and the associated loss of structural entropy within the platform. The analysis highlights that the integration of new users enriches the structural complexity of network. These transactional activities of users generate discernible patterns of behavior that are amenable to tracking and analysis. It is inferred that the extent of privacy compromise, quantified by the reduction in structural entropy, increases in tandem with the growth of the new user base on the platform.

The finding that link privacy leakage escalates among users who consistently participate on the platform is noteworthy. This pattern suggests that continuous interaction may unintentionally result in a progressive reduction of privacy. This is because sustained engagement often creates new connections or strengthens existing ones in the network. Especially when these connections are tenuous or fragile, they might unintentionally expose users to increased privacy threats, rendering their transactional activities more susceptible to scrutiny and tracking.

Conversely, the study indicates that users who reduce their social interactions on the platform, thereby becoming less active, may nevertheless face link privacy risks. This susceptibility arises from the intricate web of enduring connections within the social network. The behavior of persistently active users can unveil patterns that inadvertently implicate less active individuals, especially if past engagements have interconnected them within the transactional history.
Furthermore, the interactions of active users with layer 2 accounts on disparate applications add complexity to the privacy scenario. This points to a potential broadening of privacy concerns across multiple applications, wherein activities on one platform (other applications) may affect the anonymity on another (friend.tech). Our analysis suggests that transactional network privacy is not compartmentalized but rather interconnected across diverse applications and user activities.

By leveraging this unique intertwining of Web3 and established social media platforms, we aim to uncover the implications for user anonymity and data security. 
Our research delves into the nuances of how these interconnected identities can lead to potential privacy vulnerabilities, offering a pioneering analysis of the emergent risks in the Web3 social ecosystems.

\subsection{Recommendations} To mitigate link privacy leakage in Web3 social platforms, we suggest a dual strategy. {\itshape (i)} We recommend future Web3 platforms avoid integration with Web2 social networks. While the vast user base of Web2 is enticing, such associations may compromise blockchain user anonymity. {\itshape (ii)} We advocate for obstructing transaction linkability through advanced privacy protocols, notably Zero Knowledge Proofs (ZKPs) \cite{6956581}, to obscure the connection between users' layer 1 and layer 2 activities without undermining transaction integrity. We also encourage the adoption of Decentralized Identity (DID) systems \cite{ernstberger2023sok} to bolster user control over personal data and offer privacy-preserving authentication. These measures collectively may form a robust approach to addressing privacy issues in the Web3 social ecosystem. Our future work will explore these directions in depth. 

\subsection{Limitations} Recognizing the swift evolution of user behavior and the dynamics of platforms that could significantly influence privacy and user engagement, our study serves as a critical alert to the possible privacy breaches on Web3 social platforms. The transfer of users' funds through mechanisms such as centralized exchanges (CEXs), mixers, cross-chain bridges, or laundering services introduces a layer of complexity. These processes not only enhance user anonymity but also involve a series of intricate transactions that extend beyond the scope of our comprehensive analysis. Consequently, it is vital to acknowledge the limitations of our investigation within these complex and continually evolving systems. Moreover, the interplay between these elements underscores the need for ongoing research to adapt to and address the emerging challenges in safeguarding privacy within the rapidly growing Web3 ecosystem.

\section{Related Work} 

\textbf{Decentralised Social Networks.} Previous researches have thoroughly examined the facets of online social networks, such as structural design \cite{ahn2007analysis,cheng2008statistics,leskovec2008planetary,myers2014information} and administrative operations \cite{cao2021understanding,ganesh2020countering,geiger2016bot,gillespie2018custodians,haimson2021disproportionate,langvardt2017regulating,rice2016online,veglis2014moderation}. Some studies have probed into the foundational architectural elements \cite{kjernsmo2016sparql,schwittmann2013sonet,sharma2012supernova}, governance protocols \cite{bortoli2011decentralised}, and privacy and security challenges \cite{taheri2015security}, in addition to the aesthetics of user interfaces \cite{fisher2014designing}. Recent analyses have been directed at federated decentralized social networks \cite{bin2022toxicity,doan2020measuring,hassan2021impact,hassan2021exploring,la2021understanding,raman2019challenges}, with some researches focusing on the decentralised social networks, such as memo.cash \cite{zuo2023set}. To our current understanding, this study stands as the pioneering exploration into the investigation of link privacy within the realm of Web3 social ecosystems.
\textbf{The Transactions Deanonymity of Blockchain.} Most current researches detect the malicious activities such as money laundering, fraud, and Ponzi schemes in the use of blockchain digital currencies with transaction graphs \cite{mclaughlin2023large,li2023demystifying,das2022understanding,cernera2023token}. Dey {\itshape et al.}\cite{33} propose utilizing smart software agents to monitor stakeholder activities on blockchain, particularly against network threats in consortium chains, and to detect collusive behaviors. Pham {\itshape et al.}\cite{34} construct network graphs for users and transactions based on Bitcoin transaction data, extracting various relational features and identifying anomalies by calculating local outlier factors with density and distribution laws, along with the K-means clustering algorithm. Chen {\itshape et al.}\cite{35} employ graph analysis methods to characterize the three primary activities on Ethereum: transferring funds, creating smart contracts, and invoking smart contracts. Zheng {\itshape et al.}\cite{37} extract features from the accounts and operation codes of Ethereum smart contracts, applying data mining and machine learning techniques to uncover potential Ponzi scheme activities. Gudgeon {\itshape et al.} \cite{10.1007/978-3-030-51280-4_12} summarized the anonymity problem and challenges in blockchain, which inspired our efforts to analyze the privacy leakage between layer 1 and layer 2.



\section{Conclusion}
In this study, we delve into the burgeoning Web3 social ecosystems, with a focus on the friend.tech platform, which integrates with Twitter. Our goal is to understand how user activities on layer 2 platforms impact the anonymity of layer 1 accounts, exploring the effects of application-level interactions on blockchain identity privacy.
Our findings highlight the presence of KOLs who exhibit sustained activity, while average user participation wanes over time due to inadequate incentives, challenging the platform's decentralized ethos. Additionally, practices like wash trading and "bonus hunting" suggest an artificial popularity among users.
Our analysis reveals that active participation on layer 2 platforms can compromise the anonymity of layer 1 accounts, with even dormant users facing privacy risks. This highlights the broader privacy challenges within Web3 social networks. Our research offers insights for both academics and practitioners in the Web3 domain, emphasizing the need for strategies to safeguard user privacy. Future work will focus on developing effective privacy protection measures.



\bibliographystyle{ACM-Reference-Format}
\bibliography{references}


\end{document}